\documentclass[epj]{webofc}
\usepackage[varg]{txfonts}   
\wocname{EPJ Web of Conferences}
\woctitle{CONF12}
\begin{document}
\selectlanguage{english}
\title{Event Shape Sorting: selecting events with similar evolution}
%
%

\author{Boris Tom\'a\v{s}ik\inst{1,2}\fnsep\thanks{\email{boris.tomasik@cern.ch}} \and
        Renata Kope\v{c}n\'a\inst{2,3} 
}

\institute{Univerzita Mateja Bela, Bansk\'a Bystrica, Slovakia
\and
           FNSPE, Czech Technical University, Prague 1, Czech Republic
\and
           Rupprecht-Karls-Universit\"at Heidelberg, Heidelberg, Germany
}

\abstract{%
We present novel method for the organisation of events. The method is based on comparing 
event-by-event histograms of a chosen quantity $Q$ that is measured for each particle in every event. 
The events are organised in such a way that those with similar shape of the $Q$-histograms end-up 
placed close to each other. We apply the method on histograms of azimuthal angle of the produced 
hadrons in ultrarelativsitic nuclear collisions. By selecting events with similar azimuthal shape of their 
hadron distribution one chooses events which are likely that they underwent similar evolution from 
the initial state to the freeze-out. Such events can more easily be compared to theoretical simulations 
where all conditions can be controlled. We illustrate the method on data simulated by the AMPT model.
}
\maketitle
\section{Introduction}
\label{s:intro}

In ultrarelativistic heavy-ion collisions, the distribution of the produced hadrons reflects the dynamical 
state of the fireball at the time of its breakup. There are anisotropies of the hadron distributions which 
change from one event to another. This is a clear indication of the differences in transverse expansion
anisotropies which result from different initial conditions in each event. In many data analyses one takes 
averages over a large number of events. Naturally, questions then appear which ask about what feature
of the fireball(s) can be deduced from such averaged data. Moreover, theoretical simulations must also 
follow the same analysis procedure and so they usually must provide a large number of events which are then 
averaged over. 

We propose a more exclusive approach to data analysis. The method of Event Shape Sorting
\cite{renca} allows to select events which are similar in all details, or at least as many of them as possible. 
Hence, various higher-order anisotropies remain visible in  such data samples and are not washed away by 
averaging over different events. As a result, they can also be more directly compared to the results of 
simulations which can refrain from the need to average over many events. 

An approach in the same spirit has been used for a while now under the name Event Shape Engineering 
\cite{ESE}. There, the selection of events is made based on a single variable, e.g.\ the flow vector $q_2$,
which characterises certain selected features of each event. Our method, on the other hand, looks at the 
whole shape of the event and takes into account as many details of it as possible. 

In this short proceedings contribution we explain how the method works and provide some examples of
its application.

\section{The algorithm} 
\label{s:algo}

The method that we will apply here has been proposed and used earlier for statistical analysis
of a different kind of data \cite{andy,syne}. Each event will be represented by its histogram of the distribution 
that we focus on. Here, we shall look at distributions in the azimuthal angle, but note that an application
to rapidity distributions of the produced hadrons has also been proposed \cite{NICAwhite}.

In short, the method uses an iterative algorithm in order to sort events in a special way. After the 
sorting one arrives at an ordered sequence of events. In that sequence,  events which are close 
to each other have histograms with similar shapes, while events which end up far away from 
each other look rather different from each other. 

There is a slight complication when dealing with histograms of the azimuthal angle distribution
of the produced hadrons. They can be arbitrarily rotated. Thus, as the first step even before 
the commencement of the iterative algorithm one must choose the way how each event should 
be rotated and then perform this rotation. This step is skipped if the algorithm is used on 
distributions that cannot be rotated, e.g.~rapidity. 

Then, the algorithm goes as follows: 
\begin{enumerate}
\item (Rotate events appropriately, as discussed in the previous paragraph.)
\item Sort the events in a way that you consider reasonable. In fact, the final sorting 
does not depend on this initial sorting. Nevertheless, if the initial sorting is done 
wisely, it may decrease the number of iterations that the algorithm needs to perform. 
\item Divide the sorted events into quantiles. Again, unless the quantiles are too large, their number should 
not matter much for the final results, but it does influence the performance of the algorithm. 
We will use deciles. 
\item Determine average histogram in each quantile. 
\item For each event, characterised by its histogram entries $\{ n_i \}$, 
calculate the conditional probability $P(\mu|\{n_i\})$ that it belongs to quantile $\mu$
\begin{equation}
\label{e:bprob}
P(\mu|\{n_i\}) = \frac{\prod_i P(i|\mu)^{n_i} P(\mu)}{\sum_\nu \prod_i P(i|\nu)^{n_i} P(\nu)}
\,   .
\end{equation}
The formula has been derived in \cite{renca} and is based on Bayes' theorem. Here, $P(i|\mu)$
is the probability that in the quantile $\mu$ a random hadron would be found in angular bin $i$. 
It is calculated by dividing the number $n_{\mu,i}$ of those particles from all events in the quantile $\mu$
that fall into angular bin $i$ by the total number of all hadrons from all events in quantile $\mu$, $M_\mu$
\begin{equation}
P(i|\mu) = \frac{n_{\mu,i}}{M_\mu}\,  .
\end{equation} 
In formula (\ref{e:bprob}), $n_i$ is the actual number of particles in the angular bin $i$, measured in the 
event with the complete record $\{n_i\}$. The priors $P(\mu)$ and $P(\nu)$ are 1/10 in case of using 
deciles. 

This step is illustrated in Fig.~\ref{f:compare}. 
\begin{figure}[ht]
\centering
\includegraphics[width=0.85\textwidth,clip]{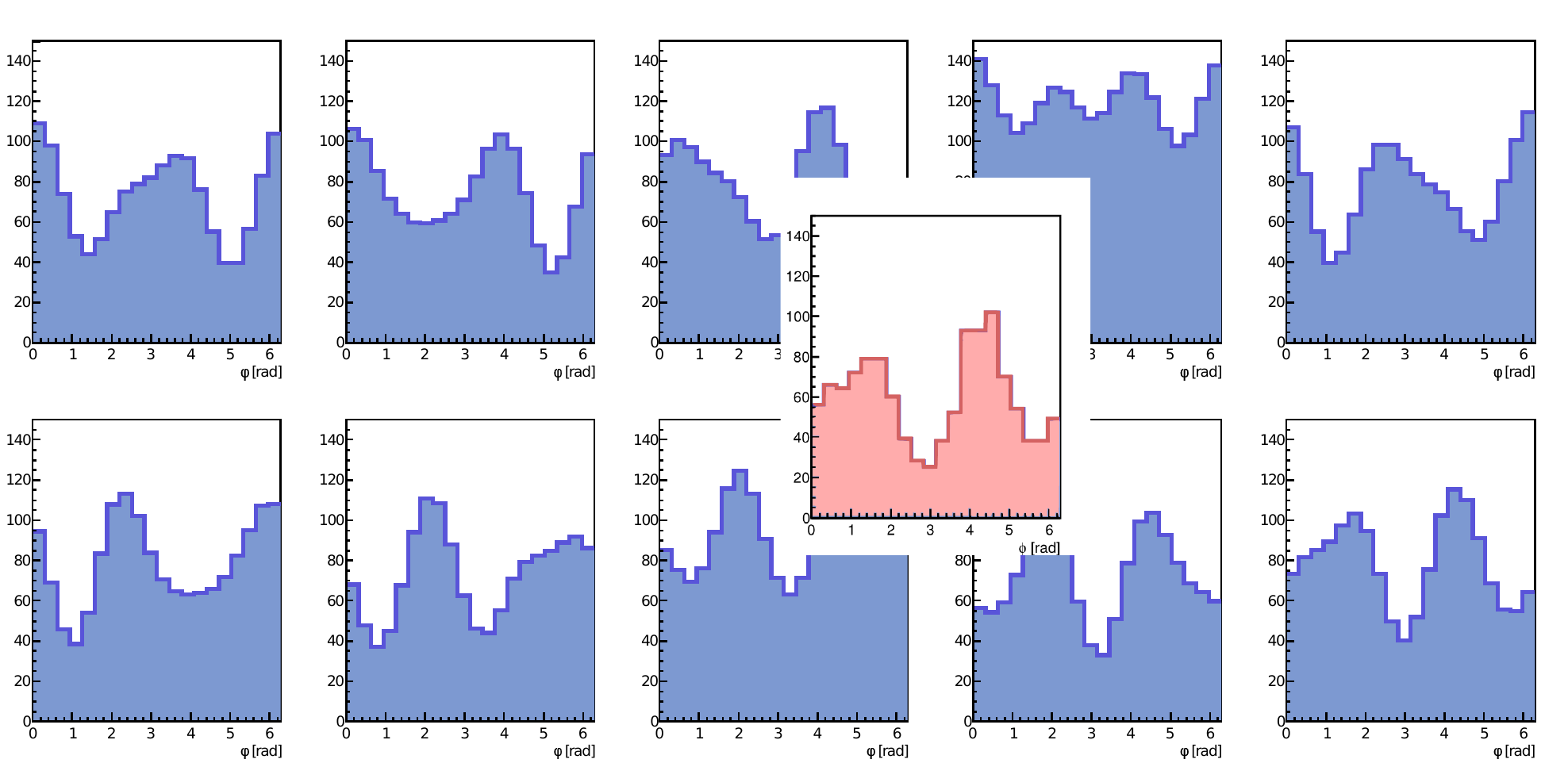}
\caption{Comparison of the histogram of azimuthal angles from a single event with 
average histograms from all quantiles.}
\label{f:compare}       
\end{figure}
The calculated probabilities basically give answer to the question: to which quantile 
(blue) looks the selected event (red) similar? Among all averaged histograms in the figure,
the selected event seems to be most similar to the last quantile, while it is certainly 
different from the fourth quantile in the upper row.
\item For each event calculate the average
\begin{equation}
\bar \mu = \sum_{\mu} \mu P(\mu | \{ n_i\})\, .
\end{equation}
\item Sort all events according to their values of $\bar \mu$. 
\item If the order of events has changed, return to point 3. Otherwise the algorithm converged.
\end{enumerate}

\section{Results}
\label{s:res}

Let us first present the action of the sorting algorithm on a trivial example, where we can also test 
that it works as expected. We will analyse artificial Monte Carlo events with only the elliptic anisotropy 
of hadron distributions taken into account. We have generated 5000 events with only pions and their 
multiplicity has been varied between 300 and 3000. The distribution of azimuthal angles followed 
\begin{equation}
P_2(\phi) = \frac{1}{2\pi} \left ( 1 + 2v_2 \cos \left ( 2(\phi - \psi_2)   \right )\right )\,  .
\end{equation}
(There was a tiny first-order term, which can be neglected in our discussion.) The second order term 
is correlated with the multiplicity. Its correlated part is determined as
\begin{equation}
\bar v_2 = a M^2 + bM + c\,  .
\end{equation}
where $a = -7.099 \times 10^{-8}$, $b = 20.06\times 10^{-5}$, and $c = 0.07874$.
The $v_2$ actually used in the event is then generated for each event as
\begin{equation}
v_2 = \bar v_2 + \tilde v_2\,  ,
\end{equation}
where $\tilde v_2$  is a random addition determined from Gaussian distribution with the width of 0.25.

\begin{figure}[ht]
\centering
\includegraphics[width=0.78\textwidth,clip]{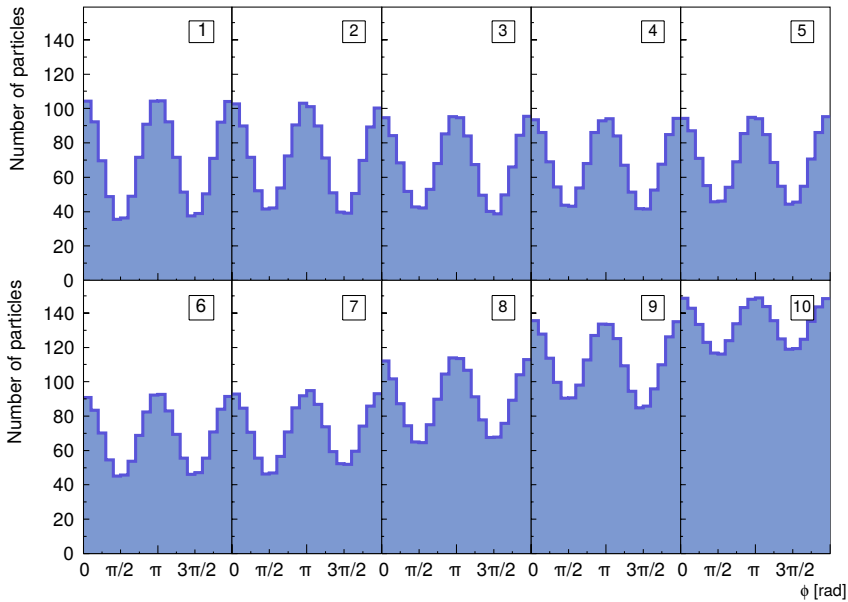}
\caption{Average histograms of ten deciles with toy-model events.}
\label{f:toy}       
\end{figure}

When  the sorting algorithm is used with such a set of events, they come out ordered according to their value
of $v_2$. This is illustrated in Fig.~\ref{f:toy}.
%
%
By inspecting the average histograms in the ten deciles we see that the elliptic variation decreases as one
moves from decile 1 towards decile 10. We have also observed that for different initial sorting, the final sorting 
may come out exactly reversed. This preserves the feature that similar events end up close to each other. 

\begin{figure}[ht]
\centering
\includegraphics[width=0.78\textwidth,clip]{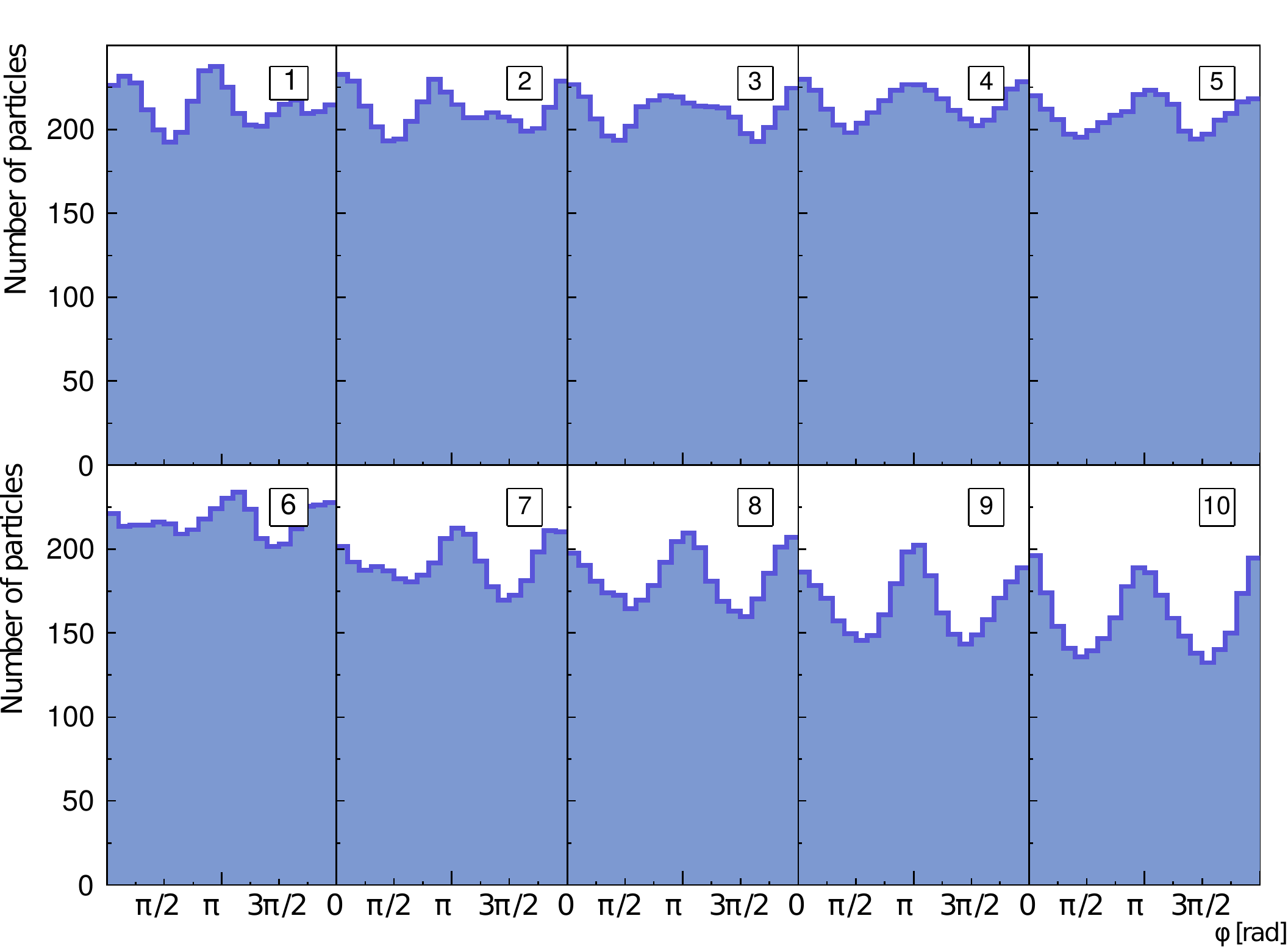}
\caption{Average histograms of ten deciles with AMPT events.}
\label{f:ampt}       
\end{figure}

We now proceed to the analysis of  events generated by the AMPT transport model \cite{AMPT}. 
In Fig.~\ref{f:ampt} we show the final average histograms of the deciles which result from 2000 events
of Pb+Pb collisions at $\sqrt{s_{NN}} = 2.76$~TeV and 0--20\% centrality class.   
%
%
Before the sorting iterations were started, events had been aligned according to their second-order
event planes $\psi_2$.

\begin{figure}[h]
\centering
\sidecaption
\includegraphics[width=0.45\textwidth,clip]{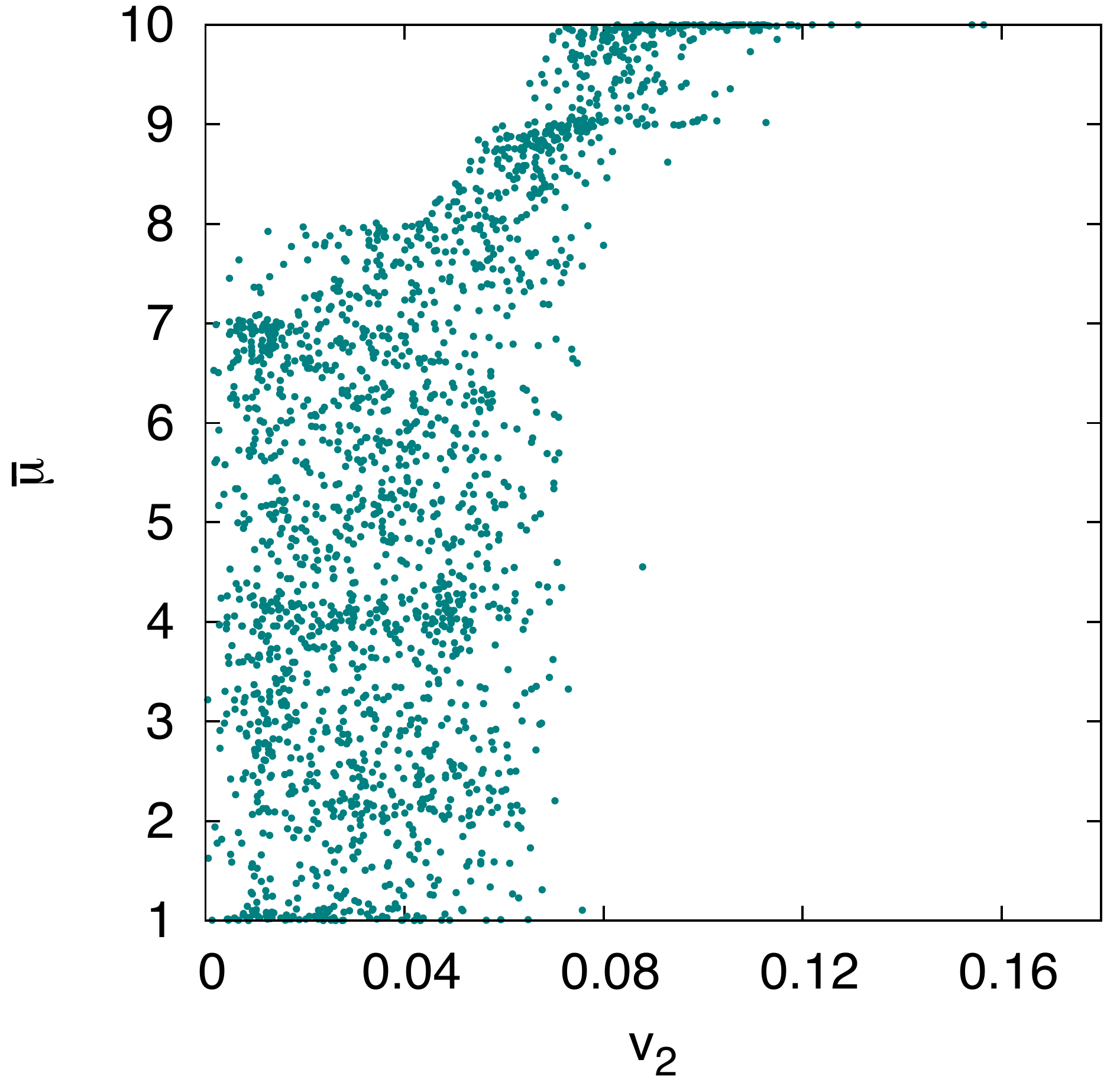}
\caption{The correlation between average $\bar\mu$ and $v_2$.}
\label{f:muv2}       
\end{figure}

It can be seen that the structure of the events is much more rich than in the previous toy model 
example. Only the deciles 7--10 show a clear dominance of the elliptic anisotropy. The rest is more 
complex. This is confirmed in Fig.~\ref{f:muv2}, where we study the correlation between the sorting 
variable $\bar \mu$ and the elliptic flow coefficient $v_2$.  
%
%
The correlation is pronounced for $\bar \mu$ above 8. For lower $\bar \mu$'s the ordering 
is determined also by other features of the distribution. Unfortunately, we were unable to determine 
how exactly it is constructed.

\section{Summary}
\label{s:sum}

Now we return to the contemplation about possible more exclusive selection of events which 
can then be compared to theoretical simulations. Once the sorting algorithm has converged,
one just has to take a sequence of events which are placed in subsequent positions. Note 
that the algorithm does not decide \emph{how} the events should look like. It only guarantees 
that the events within such a group are similar. In order to characterise them one would need 
to invoke other methods. 

Also note that so far we have not quantified the similarity of the events. Nevertheless, it should be
possible to do this with the help of some statistical method, e.g.\ by taking the Kolmogorov-Smirnov
measure of the distance between two histograms. 

With the selected group of events one could make comparisons to theoretical simulations where e.g.\ at 
the same time second and third order anisotropies would be present at fixed relative angles between 
$\psi_2$ and $\psi_3$. This is just one example of possible more exclusive studies. 

The more exclusive selection of events might also be useful for just technical purpose. 
In correlation studies, one often needs reference distributions where correlation is turned off. 
They are usually constructed by making use of the event-mixing technique: artificial events are generated 
with each particle chosen from a different event. The events from which particles are selected, 
however, must not be too different. By choosing particles from similar events determined by our 
algorithm one would minimise any unwanted influence on the constructed correlation function 
due to improper construction of the mixed-events background. 

Finally, let us speculate that if this idea is pushed to the extreme, it might open the way towards 
single-event femtoscopy. One of the problems in femtoscopy is the determination of the 
background with the mixed-events technique. This becomes critical if studies are to be done 
with single events. Nevertheless, if our method would allow to select events which are very 
similar or even identical, we might proceed with the background obtained by mixing such events
and the signal from a single event. It will be interesting to work out this speculation in the future. 


\paragraph{Acknowledgment}
BT acknowledges partial support from VEGA 1/0469/15 (Slovakia) and LG15001 (Czech Republic).
RK acknowledges support from SGS15/093/OHK4/1T/14 (Czech Republic).


\end{document}